\documentclass[preprint2]{aastex}

\newcommand{\simgt}{\,\rlap{\lower 3.5 pt \hbox{$\mathchar \sim$}} \raise
1pt \hbox {$>$}\,}
\newcommand{\simlt}{\,\rlap{\lower 3.5 pt \hbox{$\mathchar \sim$}} \raise
1pt \hbox {$<$}\,}

\shorttitle{DM attractor}
\shortauthors{Hansen, Juncher \& Sparre}

\begin{document}


\title{An attractor for dark matter structures}


\author{Steen H. Hansen, Diana Juncher \& Martin Sparre}
\affil{Dark Cosmology Centre, Niels Bohr Institute, University of Copenhagen,\\
Juliane Maries Vej 30, 2100 Copenhagen, Denmark}



\begin{abstract}
  Cosmological simulations of dark matter structures have identified a
  set of universal profiles, and similar characteristics have been
  seen in non-cosmological simulations. It has therefore been
  speculated whether these profiles of collisionless systems relate to
  accretion and merger history, or if there is an attractor for the
  dark matter systems.  Here we identify such a 1-dimensional
  attractor in the 3-dimensional space spanned by the 2 radial slopes
  of the density and velocity dispersion, and the velocity
  anisotropy. This attractor effectively removes one degree of freedom
  from the Jeans equation. It also allows us to speculate on a new
  fluid interpretation for the Jeans equation, with an effective
  polytropic index for the dark matter particles between $1/2$ and
  $3/4$. If this attractor solution holds for other collisionless
  structures, then it may hold the key to break the mass-anisotropy
  degeneracy, which presently prevents us from measuring the mass
  profiles in dwarf galaxies uniquely.
\end{abstract}




\section{Introduction}
Cosmological dark matter structures which have been formed in an
expanding universe appear to have virtually universal profiles.  This
was first emphasized for the density profile through numerical
simulation when \cite{nfw} demonstrated that all structures from
galaxies to galaxy clusters are well fitted by the same simple form
\begin{equation}
\rho = \frac{\rho _0}{\frac{r}{r_s} \left( 1  + \frac{r}{r_s} \right) ^2} \, ,
\label{eq:nfw}
\end{equation}
where $r_s$ is the scale radius.
Even though various theoretical
attempts have been made to explain such density profile
(e.g. \cite{henriksen,gsmh,hekang,hjortwilliams}) 
there is still
no concensus of the underlying physical reason for this universality.

To ease the task of theoreticians one should look for simple connections
in some para\-meter space. One such connection was suggested by
\cite{taylornavarro}, who presented numerical evidence that
the pseudo phase-space density is a power-law in radius
\begin{equation}
\frac{\rho}{\sigma^3} \sim r^{-\alpha} \, ,
\label{eq:nt}
\end{equation}
where $\alpha = 1.875$ is in agreement with the predictions from the
spherical infall model \citep{bertschinger}.  This numerical evidence
was soon used in conjunction with the Jeans equation to derive density
profiles in general agreement with eq.~(\ref{eq:nfw})
\citep{hansenjeans, austin, dehnenmclaughlin}.  One problem does,
however, present itself, namely that recent high resolution
simulations do not confirm that the pseudo phase-space density is a
universal straight line (for recent discussions, see \cite{schmidt09,
  knollmann, ludlow}).

In this Letter we study the evolution of pure dark matter structures
using numerical simulations. We find that all the various equilibrium
dark matter structures, when perturbed in a simple but realistic
manner, move towards a particular curve in the 3-dimensional space of
the derivatives of density and velocity dispersion, and the velocity
anisotropy.

\section{Structures in equilibrium}

One of the non-trivial aspects of cosmological dark matter structures
formed through hierachical mergers in an expanding universe is
that they are seldom in perfect equilibrium: the first moment
of the collisionless Boltzmann equation, the Jeans equation,
is not always exactly obeyed unless all velocity terms are included. 
When bulk motion can be ignored, this Jeans equation reads (see e.g. 
\cite{binney82})
\begin{equation}
v_c^2 = - \sigma^2_r \, \left(
\gamma + \kappa + 2 \beta \right) \, ,
\label{eq:jeans}
\end{equation}
where we have defined the derivatives of the density and radial velocity
dispersion, 
\begin{equation}
\gamma \equiv \frac{d{\rm log} \rho}{d {\rm log} r}
\, \, \, \, \, {\rm and } \, \, \, \, \,
\kappa \equiv \frac{d{\rm log} \sigma_r^2}{d {\rm log} r} \, ,
\end{equation}
the velocity dispersion anisotropy is given by
\begin{equation}
\beta \equiv 1 - \frac{\sigma_{\rm tan}^2}{\sigma^2_{\rm rad}} \, ,
\label{eq:beta}
\end{equation}
and the circular velocity is
\begin{equation}
v_c^2 = \frac{GM_{\rm tot}}{r} \, .
\end{equation}
Numerical simulations of galaxies or clusters of galaxies often have
differences between the left and right hand sides of
eq.~(\ref{eq:jeans}), of the order $20\%$ for large parts of the
structures.

This problem can be avoided by manually setting up structures in
equilibrium. There are several ways of setting up structures which
obey the Jeans equation. The first is the Eddington method, which for
a given assumed density profile creates a unique ergodic structure
i.e. depending only on the energy of the particles, and hence having
$\beta=0$ by definition \citep{eddington,binneyeddington}. This method
gives the exact distribution function, $f(\mathcal{E})$, and is
therefore often stable when evolved by an N-body simulation. When we
discuss stable configurations, we intend structures whose profiles,
$\rho(r), \sigma_r^2$ and $\beta(r)$, remain unchanged when evolved for
several dynamical times by an N-body code.

The other method is to make an assumption for the radial profiles of
the density and velocity anisitropy, and then solve the Jeans equation
to get $\sigma_r^2$ \citep{hernquist93}. In this method the velocity
distribution function is assumed to be a Gaussian, and the system will
therefore not be exactly stable when evolved by an N-body simulation.
However, after having run for a number of dynamical times, such
structures settle down in an equilibrium configuration, which often
has a fair resemblance to the original density and velocity anisotropy
profiles. Below we will use both methods of defining initial
conditions.

\section{Flow towards an attractor}

The Jeans equation in eq.~(\ref{eq:jeans}) has infinitely many
solution: there are 3 unknown functions, $\rho(r), \sigma_r^2 (r)$ and
$\beta (r)$, and only one equation.  It is therefore no mystery that
one can create a wide range of different structures, with different
density and velocity profiles, which all obey the Jeans equation. The
question we are posing here is, if all these structures, if perturbed
in a simple way, will flow to new points in a large volume in solution
space, or if there is one or several attractors which the structure
will be drawn towards.

As solution space it is natural to consider the 4 dimensional space
spanned by $r, \gamma, \kappa$ and $\beta$, see
eq.~(\ref{eq:jeans}). Whereas the Jeans equation allows solutions to
fill a large 3-dimensional volume, then {\em if} an attractor does
exists, we should expect all final structures to land on a low
dimensional sheet, potentially a line. In practice we divide each
structure into radial bins. For each radial bin we can derive $\gamma,
\kappa$ and $\beta$, and this radial bin thus represents a point in 3d
space.  In this way we also avoid any potential problems concerning
scaling different structures in radius.

During the formation of cosmological structures in the real universe,
these are continuously being perturbed through the process of mergers.
This implies that {\em if} an attractor exists in solution space, then
we might expect repeated perturbations, such as the ones from ongoing
mergers, to support the flow of the structures towards the attractor.

In order to mimic the perturbations of the mergers in a controlled
way, we perturb the equilibrated structure in a simple yet fairly
realistic manner.  The velocity of each particle is multiplied by a
random number, in such a way that any given radial bin has exact
conservation of kinetic energy and angular momentum. We assure that no
particles get a velocity above the escape velocity.  Naturally, the
potential energy is also conserved during the perturbations.  Now,
having perturbed each particle in the structure in this controlled
manner, we start the N-body code, and let the structure settle into a
new equilibrium condition. We let each simulation run for a time
corresponding to one dynamical time, $t_{\rm dyn} = 1/\sqrt{G\rho}$,
at radius 13 times the scale radius.

The simulations were performed with the public version of Gadget-2
\citep{springel2005,springel2001} which is a massively parallel N-body
code that uses a hierarchical tree algorithm to calculate
gravitational forces, and gives individual timesteps to all
particles. A spline softening of $0.005 r_s$ was used. 
We are being overly conservative in the ana\-lysis, and include only
regions outside of 5-10 times this softening. We also require that
the bins contain at least $5000$ particles.

\section{The wide range of initial conditions}

\begin{figure}[thb]
        \centering
        \includegraphics[angle=0,width=0.49\textwidth]{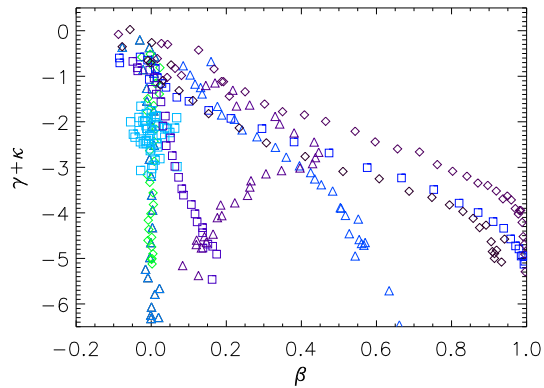}
        \caption{The initial conditions. A 2-dimensional projection in
          solution space to the Jeans equation.  According to the
          Jeans equation, eq.~(\ref{eq:jeans}), all points in this
          space may be populated.  No stable solutions exist in the
          upper right corner. Our various initial conditions of
          equilibrated structure are created to span a large volume in
          the 3-dim solution space. All the structures of
          this figure are in equilibrium, and are stable when evolved
          in an N-body simulation.}
\label{fig:ic}
\end{figure}

In order to allow the systems to start from a large range of initial
conditions, we initiate a set of significantly different
systems. Using the Eddington method, we have an isotropic system,
which has a density profile like a Hernquist
profile~\citep{hernquist90}, with inner and outer density slopes of
$-1$ and $-4$. We construct a cored profile, with central slope of
$0$, and outer slope of $-5$. We also use a {\em hoovering} profile,
$\rho \sim 1/(1+(r/r_1)^2 \,(1+r/r_2)^3)$, where $r_1=0.01$ and
$r_2=100$ which is characterized by having a large part of the central
structure where the slope is $-2$. We also create a set of initial
conditions with Osipkov-Merrit anisotropies \citep{osipkov,merritt}
with the property $\beta (r) = 1/(1 + (r_a/r)^2)$: a Hernquist density
profile (slope $-1, -4$) with $r_a=1$, and a cored profile (slope
$0,-3.5$) with $r_a =1.5$. Using the gaussian method, we create 4
structures, with both trivial and highly non-trivial $\beta (r)$
profiles, all with a Hernquist density profile (slope $-1, -4$). All
structures presented here contain $10^6$ DM particles, and low
resolution runs with $10^5$ particles establish that this is
sufficient for our purpose.  All of these initial conditions (after
having run for 3 dynamical time at radius 13 times the scale radius)
are presented in figure \ref{fig:ic}.  This figure shows that our
initial conditions span a large part of the solution space of ($\beta,
\gamma + \kappa$). The Jeans equation allows all of this parameter
space to be populated, however, there are no initial conditions in the
upper right corner, where the inequality $\beta < -\gamma/2$
\citep{anevans} prevents any stable solutions (see also
\cite{ciotti}).

\section{The perturbations}

All perturbations follow the repeated pattern of {\em kick-flow}.
First the structure is perturbed (kicked) and then the N-body
simulation allows it to settle into (flow towards) a new equlibrium
configuration. This kick-flow is repeated numerous times.

We first consider 2 structures, each with initial Hernquist profile
and with $\beta=0$. These are created using the Eddington and the
Gaussian method respective. We then perturb the energies of each
individual particle in the range $[ 0.75, 1.25 ]$, in such a way that
there is energy conservation at each radial bin.  The structure is
then evolved with Gadget-2 for 1 to 3 dynamical time at 13 times the
scale radius.  This process is repeated approximately 20 times. The
final structures are compared, and we find no significant difference
when plotted in the plane ($\beta, \gamma + \kappa$). This shows that
there is virtually no dependence in the final result on the method of
creating the initial conditions (Eddington vs. Gaussian).

We then take the isotropic Hernquist structure, using the Eddington
method, and perturb by changing energies in the much larger range $[
0.25, 1.75 ]$. Also this structure ends in the same region in ($\beta,
\gamma + \kappa$) after repeated perturbations. This demonstrates that
the magnitude of the perturbation is not crucial to the final
structure.  We also test perturbations where we don't prevent
velocities above the escape velocity: for these structures some
particles do leave the structure, and these perturbations are
therefore less conservative. Also structures perturbed in this way
land on the same curve in solution space.  We can therefore perturb
all structures by changing energies in the range $[ 0.25, 1.75 ]$,
irrespective of the method of creating the initial conditions.

\begin{figure}[thb]
        \centering
        \includegraphics[angle=0,width=0.49\textwidth]{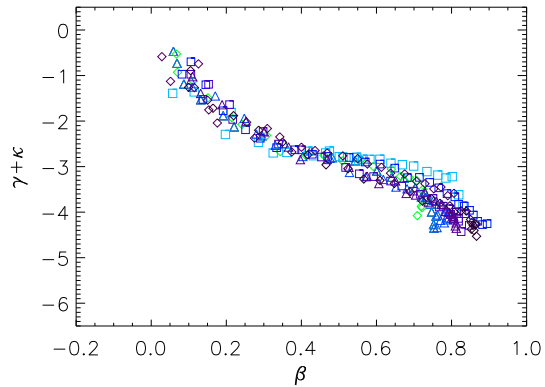}
        \caption{The attractor solution. The same 2-dimensional
          projection in solution space of the Jeans equation,
          eq.~(\ref{eq:jeans}), as presented in Fig.~1. The points
          represent the final profiles for all the initial
          conditions, after these have been perturbed and allowed to
          equilibrate repeatedly. This attractor solution is also a
          1-dim curve in the full 3-dim space spanned by $\gamma,
          \kappa$ and $\beta$.}
\label{fig:final}
\end{figure}

The result of all these perturbations is presented in figure
\ref{fig:final}, from which it is clear, that all the structures end
up along a 1-dimensional curve in this 2-dim parameter space. It is
important to emphasize that we had initial conditions both above and
below this attractor solution. When we plot in the 3-dim space spanned
by $\gamma, \kappa, \beta$ we still find that the attractor solution
spans a non-trivial 1-dim curve.

This curve is clearly S-shaped, but we can approximate it with a
piecewise straight line, which for small $\beta$ could be $ (\gamma +
\kappa) = - 8\beta$, and for intermediate $\beta$ it looks like
$(\gamma + \kappa) = -0.7 - 4 \beta$. It is interesting to note that
the anisotropy-slope connection suggested by \cite{hansenmoore} looks
like $(\gamma) = -0.8 - 5 \beta$ in the region which can be trusted,
namely $-2.5 < \gamma < -1$
\citep{hansenstadel,zait,2010MNRAS.402...21N}, which is in good
agreement with our findings.
For very high $\beta$ we are in the very outer region of the
structures, where full equilibrium can be questioned, since a larger
spread in the final points is observed. The included points have been
allowed to equlibrate at least 20 dynamical times in total, however,
the outermost points have only been allowed 1 dynamical times since
the last perturbation.  We therefore won't speculate on the form of
the attractor there.

Whereas the 1-dimensional attractor has a non-trivial S-shape in the
space of $\gamma, \kappa$ and $\beta$, then it happens to lie in a
simply parametrized sheet, given by one single free parameter,
$\delta=0.15$,
\begin{equation}
\beta = \frac{Y}{ \left( 1 + Y^3 \right) ^ {\frac{1}{3}}  }
\label{eq:fit}
\end{equation}
where $Y=-\delta \gamma - (1-\delta) \kappa$. This means that when
seen in a slightly different projection, the attractor in
fig.~(\ref{fig:final}) looks like a very simple curve.

With this attractor we can effectively remove one degree of freedom
from the Jeans equation: for a given density profile there is now a
unique connection between $\kappa$ and $\beta$, which allows us to
calculate both.

\section{An effective equation of state}

The Jeans equation describing collisionless particles is fundamentally
different from the Euler equation for collisional fluids, however, a
standard mapping between the two is made by setting $\beta=0$ and
interpreting $m \sigma^2_r/2$ as a temperature
\citep{binneytremaine}.  The non-trivial part of this interpretations
is to assume that $\beta =0$, since numerical simulations for years
have demonstrated that $\beta$ is not generally zero
\citep{carlberg,colelacey}.

We now allow ourselves to use our newly found connection to make a
different interpretation of the Jeans equation from a fluid
perspective.

Let us first consider the region with small $\beta$, where the
attractor solution approximately reads $ (\gamma + \kappa) =
- 8\beta$. When we insert this into the Jeans equation, we get
\begin{equation}
\frac{GM_{\rm tot}}{r} = - \sigma^2_r \, A \, \left(
\gamma + \kappa  \right) \, ,
\label{eq:jeans2}
\end{equation}
where $A=3/4$.  If we attempt to describe the dark matter by an
effective equation of state (generalizing a polytropic index) we have
\begin{equation}
P = b \, \left( \rho \sigma_r^2  \right)^{K_0} \, ,
\label{eq:eos}
\end{equation}
where $b$ is a constant.  Thus, if we require that the Jeans equation,
eq.~(\ref{eq:jeans2}), should take the form of the equation of hydrostatic
equilibrium, $GM/r^2 = \nabla P/\rho$, then we find the solution $K_0 = 3/4$. We can therefore
interpret this by the DM behaving like a fluid with an effective
polytropic index of $3/4$.

For intermediate $\beta$ we found approximately the connection
$(\gamma + \kappa) = -0.7 - 4 \beta$. This is almost solved in
the same way, finding an effective polytropic index of $1/2$,
however, the constant term ($B=-0.7$) changes the picture. In order to solve
for such an effective polytropic index, we generalize
eq.~(\ref{eq:eos}) to $P = b \, \rho^{K_1} \, \sigma_r^{2K_2}$, where
$b$ is a new constant. We now need to introduce an assumption
about the pseudo phase-space
density, namely $\rho/\sigma_r^\epsilon \sim r^{-\alpha}$. Doing this
we find the solution, $K_1 = 1/2 - B/\alpha$, and 
$K_2 = 1/2 + B \epsilon /2\alpha $.

\section{Will any perturbation do?}

A ball will always run downhill, if kicked randomly by a group of
children playing, however, if they consistently kick the ball uphill,
then the ball may appear to defy the attractiveness of gravity. We
should expect the same to be possible with the abovefound
attractor. In order to quantify this idea we tried to perturb {\em
  only} the radial velocity component. When the radial component is
perturbed by a random number (always obeying exact energy conservation
in any radial bin) then the radial velocity distribution gets flattend
out. When we subsequently let the system evolve by the N-body
simulation, the high energy tail of the distribution essentially gets
cut off, and thereby the effective radial dispersion gets slightly
smaller. When we repeatedly perform such kick-flow, we even manage to
get negative values of $\beta$ at intermediate radii. This confirms
our expectation that perturbations may be invented, which will defy
the attractor. If we instead only perturb the tangential velocities,
then the system flows towards the attractor.

We see from this discussion that it is relevant to consider which kind
of perturbations should happen during structure formation.  During
merging there will be violent relaxation. Two salient features of
violent relaxation is that it changes the energy of individual particles
on a timescale similar to the dynamical time \citep{lyndenbell67},
which for mergers is of the order $R/v$, where $R$ is the size of the
smaller merging object and $v$ is its velocity. We are here interested
in the energy exchange with specific particles in the larger
structure, which has a different timescale from the equilibration of
the smaller structure.  The other feature is that violent relaxation
acts on all particles irrespective of their orientation. In our
example above we have an instantaneous exchange of energies between
the particles (reminicent of violent relaxation) and then we let the
system flow (which provides the needed phase mixing).

In our particular setup, the {\em kicks} do not change any properties
of the profiles by themselves, and without the kicks there is no
effect of the {\em flow}. Thus, when interpreted as violent relaxation
(kick) and phase mixing (flow), we see that both are needed to move
towards the attractor.

Antonov's laws of stability state that e.g. an isotropic Hernquist
structure will remain isotropic when exposed to minor perturbations.
These laws are derived under the assumption that the r.h.s. of the
Boltzmann equation is zero. Our perturbations act like an
instantaneous non-zero term on the r.h.s. of the Boltzmann equation,
and we are therefore not violating Antonov's laws of stability.

Finally, it is worth emphasizing that we find the attractor in the
space of $\gamma, \kappa$ and $\beta$. Very specifically, we do not in
this work identify one unique density profile, $\rho (r)$.  This may be
because we are constraining the perturbations to allow very little
freedom in mo\-difying the density profile, or it may be because the
universality of the density profile \citep{nfw,moore} really finds its
origin in accretion history \citep{gsmh}.

\section{Breaking the mass velocity anisotropy degeneracy}

Let us finally discuss what our newly found attractor may do for the
mass-velocity anisotropy degeneracy. When we observe the stellar
kine\-ma\-tics in a dwarf galaxy we can observe the stellar density
and the stellar dispersion.
Then, the jeans equation, eq.~(\ref{eq:jeans}), tells
us that for any assumed velocity anisotropy profile for the stars,
$\beta(r)$, we can solve for the total gravitating
mass. However, if we had assumed a different $\beta(r)$ then we would
have found a different total mass profile \citep{strigari}.

If we for instance consider a Hernquist density profile with a
$\beta$ profile in agreement with numerical simulations and
observations \citep{hansenpiff2007,host2009,wojtak2010}, then the
reconstructed mass is overestimated by up to $40\%$, if the analysis
is made under the simplifying assumption $\beta=0$.  Also the derived
inner density slope (from the total mass) is systematically found
to be more shallow than the true slope is, by up to $10\%$. This
means that if the true density slope is $-1$, then we will measure
around $-0.95$, if we assumed $\beta =0$ in the analysis.

This may no longer have to be the case. If our attractor solutions
also applies to stellar systems in a dwarf galaxy, or to the dynamics
of the galaxies in a galaxy cluster, then we have a unique connection
between the 3 quantities, $\gamma, \kappa$ and $\beta$. Therefore, if
we have measured (accurately) the stellar density and dispersion
profiles, then we do in principle know exactly what $\beta(r)$ looks
like, and we can then deduce the unique total mass profile.

\section{Conclusions}

We have identified an attractor solution for dark matter structures.
This implies that any dark matter structure which is repeatedly
perturbed (e.g. through violent relaxtion during merging) and then
allowed to relax (phase mix), will flow towards this 1-dimensional
curve in the 3-dimensional space spanned by the 2 radial derivatives
of the density and velocity dispersion, and the velocity anisotropy.

This finding provides strong support for the idea that the
universalities found in cosmological dark matter structures are a
property of gravity, and not simply a result of
similar accretion and merger histories of different structures.

This attractor solution effectively removes one degree of freedom from
the Jeans equation, giving hope that we will eventually be able to
solve the Jeans equations analytically, and thereby truely understand
the origin of the universal profiles.

\noindent

{\bf Acknowledgements}\\ It is a pleasure to thank Jens Hjorth for discussions.
The simulations were performed on the facilities provided
by the Danish Center for Scientific Computing.
The Dark Cosmology Centre is funded by the Danish National Research Foundation.



\end{document}